\documentclass{iau}

\usepackage{amsmath}
\usepackage{graphicx}
\usepackage{multirow} 

\begin{document}

\lefttitle{Benedetta Casavecchia}
\righttitle{Absorption spectra from galactic wind models}

\jnlPage{1}{7}
\jnlDoiYr{2021}
\doival{10.1017/xxxxx}

\aopheadtitle{IAU Symposium No. 362}
\editors{Dmitry Bisikalo, Tomoyuki Hanawa, Christian Boily, eds.}

\title{Absorption spectra from galactic wind models: a framework to link PLUTO simulations to TRIDENT}

\author{Benedetta Casavecchia$^{1,2}$, Wladimir E. Banda-Barrag\'an$^2$, Marcus Br\"uggen$^2$, and Fabrizio Brighenti$^1$}

\affiliation{$^1$Dipartimento di Fisica e Astronomia, Universit{\`a} di Bologna, Via Gobetti 93/2, 40122, Bologna, Italy \\ email: {\tt benedett.casavecchia@studio.unibo.it} \\
$^{2}$Hamburger Sternwarte, Universit\"{a}t Hamburg, Gojenbergsweg 112, D-21029 Hamburg, Germany}

\begin{abstract}
Galactic winds probe how feedback regulates the mass and metallicity of galaxies. Galactic winds have cold gas, which is mainly observable with absorption and emission lines. Theoretically studying how absorption lines are produced requires numerical simulations and realistic starburst UV backgrounds. We use outputs from a suite of 3D PLUTO simulations of wind-cloud interactions to first estimate column densities and temperatures. Then, to create synthetic spectra, we developed a python interface to link our PLUTO simulations to TRIDENT via the YT-package infrastructure. First we produce UV backgrounds accounting for the star formation rate of starbursts. For this purpose, we use fluxes generated by STARBURST99, which are then processed through CLOUDY to create customised ion tables. Such tables are subsequently read into TRIDENT to generate absorption spectra. We explain how the various packages and tools communicate with each other to create ion spectra consistent with spectral energy distributions of starburst systems.
\end{abstract}


\maketitle

\section{Introduction}

It is now well known how important it is to no longer consider galaxies as isolated systems. Galaxies are dynamic systems that acquire gas from the surrounding media. They can grow via merger events or by accreting material from the intergalactic medium, which then provides the fuel for star formation. Each galaxy is immediately surrounded by a galactic corona known as the circumgalactic medium (CGM), which contains gas, dust, and cosmic rays \citep{2017ARA&A..55..389T}. Observations of such material reveal that the CGM is complex because of feedback processes that allow for the exchange of matter and energy. Feedback can be due to star formation or active galactic nuclei (AGN), and it often has a strong impact on a galaxy's evolution, its star formation rate (SFR) and its metallicity.\par

In this project we solely focus on stellar feedback, which is observed in starburst galaxies, where most of the released energy comes from the SNe explosions and stellar winds from young massive OB-type stars, and manifests as galactic winds (see section 8.7 of \citealt{2019arXiv191206216C,2020A&ARv..28....2V,2021ApJ...921...91D}). For nearby galaxies, outflow patterns have been studied in emission and absorption lines, but for distant systems, observations are generally limited to the study of absorption lines due to cold gas around galaxies (e.g. \citealt{2017ARA&A..55..389T} and \citealt{2018MNRAS.474.1688C}). These are detected either down-the-barrel with the stellar continuum of the galaxy in the background or transversely, i.e., along the lines of sight of distant quasars (QSOs). From the study of absorption lines it is possible to extract relations for the estimation of the SFR and the stellar mass of star-forming galaxies, as for example was done in \cite{2021arXiv211013167T}, where the authors used the O VI column density for their estimates.\par

Simulations, as is often the case in astrophysics, play a key role in better understanding the physics behind the observed phenomena. Indeed, there are many examples of simulations that attempt to emulate the behaviour of atomic/molecular clouds interacting with galactic winds (e.g. \citealt{2015ApJ...805..158S,2018MNRAS.476.2209G,2019MNRAS.486.4526B,2021ApJ...919..112D}). Our project ultimately aims at finding a link between numerical models and observations. For the first part of our project, which we present in this paper, we have created a tool that is able to produce column density maps and absorption spectra of different ions from grid-based simulations, specifically from the PLUTO code \citep{2007ApJS..170..228M}, in order to compare simulations of galactic winds with observations.

\section{Simulations}

\begin{table}\centering
\caption{Initial conditions for single cloud models interacting with galactic winds. Column $1$ indicates the name of the model, column $2$ shows the value to which the cooling floor is set, and column $3$ indicates the orientation of the magnetic field with respect to the wind. The computational domain of the simulation is $384 \times 768 \times 384$, which corresponds to a physical domain of $(120 \times 240 \times 120)$ pc; thus, 32 cells cover the initial radius of the cloud. Initially, the cloud has a spherical shape with uniform density distribution, it is at rest with a temperature of $8.06 \times 10^2$ K and has a polytropic index $\gamma = 5/3$. The wind-cloud density contrast $\chi = \rho_{\rm cloud}/\rho_{\rm wind} \approx 700$, the wind Mach number ${{\mathcal M}}_{\rm wind}= 4$ and plasma beta parameter $\beta = P_{\rm th}/P_{\rm mag} \approx 100$; where $\beta$ is defined as the ratio of the thermal pressure to the magnetic pressure of the plasma.}
 \hspace*{-0.35cm}\begin{tabular}{c c c}
\hline
\textbf{(1)} & \textbf{(2)} & \textbf{(3)}\\
\textbf{Model} & \textbf{Cooling Floor [K]} & \textbf{Magnetic Field Orientation}\\\hline
rwc-r32-al-f1 & $50$ & Aligned\\
rwc-r32-tr-f1 & $50$ & Transverse\\
rwc-r32-al-f2 & $10^4$ & Aligned\\
rwc-r32-tr-f2 & $10^4$ & Transverse\\\hline
\end{tabular}
\label{Table1}
\end{table} 

To develop the framework presented in this paper, we have utilised data from existing simulations of wind-cloud models based on \cite{2016MNRAS.455.1309B} and \cite{2018MNRAS.473.3454B}. All the simulations presented here use a 3D Cartesian coordinate system ($X_1$, $X_2$, $X_3$) with uniformly-spaced grid cells, and resort to standard methods for ideal, grid-based hydrodynamics, available in the PLUTO code \citep{2007ApJS..170..228M}, to solve the mass, momentum, and energy conservation laws. Further details on the numerics of these models will be presented in upcoming publications.\par

Our models account for interactions between isolated spherical clouds and supersonic winds in the presence of weak magnetic fields. The gas is subjected to radiative heating and cooling, so it can lose or gain thermal energy depending on its temperature and density. During the computation, the cooling and heating rates are read from a table (preliminary produced by the CLOUDY code by \citealt{2017RMxAA..53..385F}) and then they are removed or added to the energy conservation equation (in a similar fashion to what is described in \citealt{2021MNRAS.506.5658B} for shock-multicloud models, see also \citealt{2020MNRAS.499.2173B}). The initial conditions and domain sizes of our models are displayed in Table \ref{Table1}.\par

\begin{figure}[h!]
   \centering
   \includegraphics[width=1.0\textwidth]{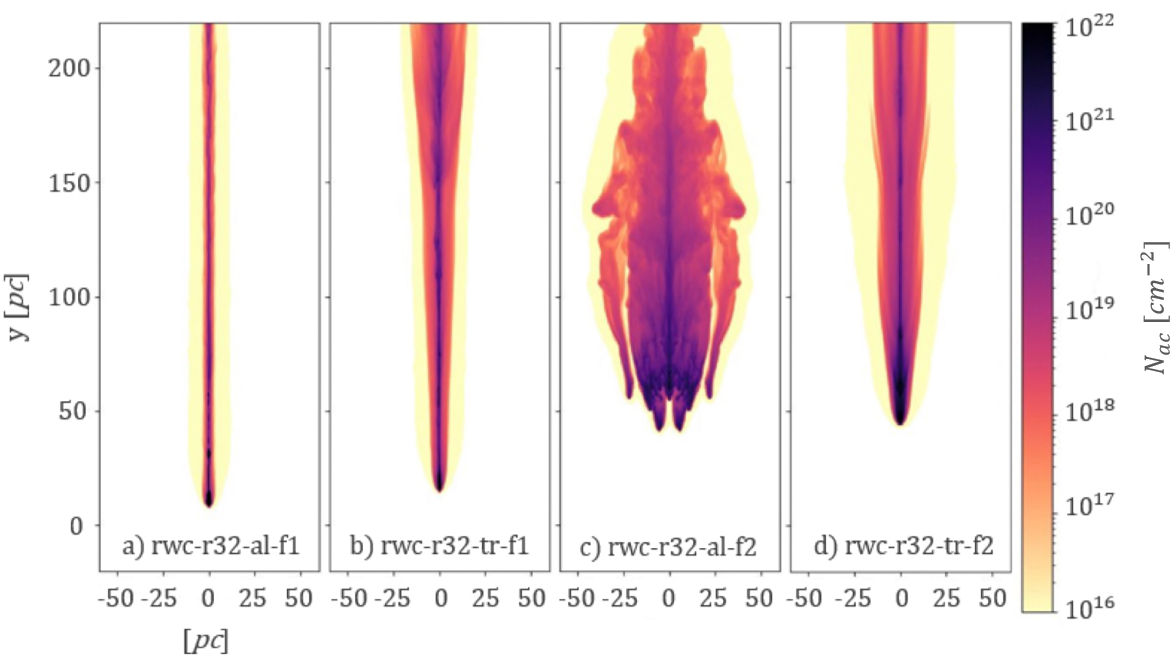}
   \caption{Column number density of the clouds integrated along the $ X_3 $ axis. The 4 panels aim to demonstrate that the choice of the cooling floor and magnetic field orientation can drastically change the evolution of wind-swept atomic clouds. The images are taken at the same time and represent the $4$ models shown in Table \ref{Table1}. They show the generation of filaments as the clouds begin to lose material when interacting with the wind. Some differences between the models persist; in panel a) and b) the clouds are not completely destroyed by the end of the simulation and their centre of mass does not move excessively from the starting point. On the other hand, in the models with the cooling floor set to $10^4$ K, shown in panels c) and d), the clouds do not have the chance to cool down beyond that value, so they expand and are therefore displaced along the wind direction. Finally, in the latter two models, the orientation of the magnetic field has a stronger influence on the cloud evolution, leading to evaporation if the magnetic field is aligned with the wind, and a slower disintegration when it is transverse.}
   \label{fig:coldens40}
\end{figure}

From a physical viewpoint, the simulations we analysed are wind-tunnel simulations representing 3D sections of a global galactic wind. In particular, they emulate the behaviour of atomic clouds interacting with supersonic flows that represent galactic winds. The main features that our four simulations have in common are mentioned in the caption of Table \ref{Table1}. All of them have been implemented taking into account the same atomic radiative processes, the same magnetic field configurations, and a uniform cloud density distribution with the same density contrast with respect to the wind. On the other hand, the differences that have been implemented in the initial conditions of our four panels include: a different cooling floor ($50$ K or $10^4$ K) and a different direction of the magnetic field (parallel to the wind direction or transverse to it). As can be seen in Figure \ref{fig:coldens40}, these variations in the initial conditions lead to very different dynamical evolutions.

\section{Generation of Synthetic Spectra}

\begin{figure}[h!]
   \centering
   \includegraphics[width=0.75\textwidth]{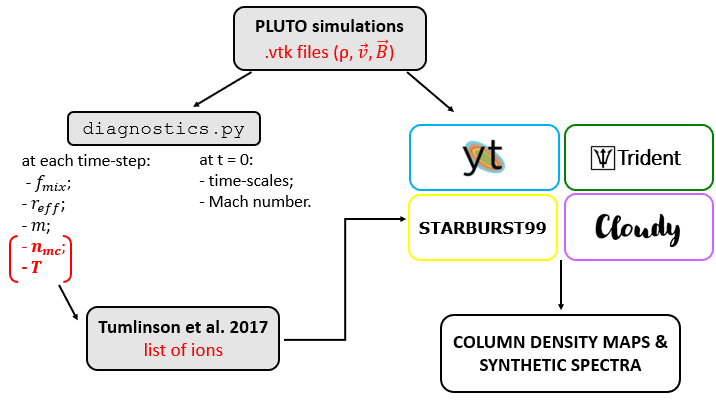}
   \caption{Flow chart summarising the main components of the framework that produces column density maps and synthetic spectra from PLUTO simulations. The starting point is the datasets provided by PLUTO, which we initially analysed using the script \texttt{diagnostics.py} in order to find which ions would be included in the synthetic spectra. The right-hand side of the diagram shows which software we used to produce the ion density maps and spectra.}
   \label{fig:flowchart}
\end{figure}

Here we discuss the implementation of a framework able to produce synthetic spectra and column density maps of atomic/molecular clouds interacting with galactic winds from the simulation dataset described in the previous section. The flowchart shown in Figure \ref{fig:flowchart} provides a summary of the framework. In the next subsections we will explain in more detail how each component of the code and the various software packages interact with each other, in particular how spectra are produced by combining two Python tool-kits, YT \citep{2011ApJS..192....9T} and TRIDENT \citep{2017ApJ...847...59H}, together with the STARBURST99 \citep{2017IAUS..316..359V} software, and the CLOUDY Code \citep{2017RMxAA..53..385F}.\par

\subsection{Preliminary analysis and data preparation}

The first part of the framework involves reading the \texttt{.vtk} files produced by the PLUTO simulations. Then, we analyse the data in order to study thermodynamical aspects of these systems. For this purpose, we have created a script in Python called \texttt{diagnostics.py} that receives as input the \texttt{.vtk} files with the gas density, velocity, magnetic field of each cell and normalisation factors in c.g.s units ($\rho_0$, $v_0$, $L_0$, plus the mean particle mass, $\mu$, and the polytropic index $\gamma$).\par

Our analysis in Python provides two types of outputs: a set of diagnostics calculated at $t = 0$ such as: magnetic field strength, initial temperature, Mach number and time-scales; and a set of time-dependent quantities averaged over the whole simulation volume and calculated at each simulation time step, such as: number density, cloud mass, coordinates of the cloud's centre of mass, its effective radius, temperature and mixing fraction. Our scripts also produce column density maps of cloud gas (both edge-on and down-the-barrel) and 2D histograms which allow us to determine which ions should exist, given the number density-temperature ranges characteristic of these interactions. Using these results, and the inputs from Figure 6 in \cite{2017ARA&A..55..389T} we can readily identify which ions should be typically observed in our simulated galactic winds, and can therefore be targeted in our research.\par

\subsection{Reading PLUTO data into the YT framework}

A more detailed picture of how the synthetic spectra are generated can be deduced from the flow chart in Figure \ref{fig:flowchart}, which shows the key steps of the process. The first package we used was YT. YT is a toolkit for visualisation and analysis of datasets produced by 3D simulations and it is also the first dependency of TRIDENT. YT has loaders for several commonly-used hydrodynamic codes, but does not have a native interpreter for PLUTO data, so it was necessary to create a script to fill this gap. The Python script in matter is called \texttt{Pluto\_YT\_interface.py}, which converts the content of each .vtk file into a series of numpy arrays. Thanks to the \texttt{yt.load\_uniform\_grid} function, it is possible to load the data and display temperature and density maps. Subsequently, this function creates a dictionary with all the data fields and their units in c.g.s. and defines a computational domain that coincides with the simulation one.\par

\subsection{Feeding TRIDENT and producing spectra}

Once the grid-based data have been read into YT as a ``StreamHandler" and once the ions to be analysed have been chosen, we produce column density maps and temperature-number density 2D histograms to compare with those produced earlier by the \texttt{diagnostics.py} script. If everything matches, the following step is to produce synthetic spectra exploiting the TRIDENT package. The first function used is \texttt{trident.add\_ion\_fields} which generates projected density maps for each ion in order to better understand their spatial distribution within the cloud.\par

The second step is to use the \texttt{trident.make\_simple\_ray} function to save all the field values of the ions intercepted by the beam and save them in an \texttt{.h5} file (see \citealt{2017ApJ...847...59H}). Finally, the last part of the code involves the \texttt{trident.SpectrumGenerator} class which receives as input the ray and the range of wavelengths in Angstrom, the resolution $\Delta \lambda$ and it generates as output the spectrum. In this final stage it is possible to decide the type and intensity of noise in order to produce a more realistic spectrum, and also to produce it in velocity space.

\subsection{Customised UV Backgrounds}

TRIDENT uses by default the \cite{2012ApJ...746..125H} UV background, which only takes into account the cosmic UV background radiation. However, since the atomic clouds investigated in this study are located close to regions of high star formation, it is necessary to consider the extra UV radiation produced by the evolution of e.g. OB stars (see also \citealt{2018ApJ...864...96C}). The creation of a more realistic background is briefly summarised in the top right-hand part of Fig. 2. We used the STARBURST99 software and the CLOUDY code, in particular the library \texttt{Cloudy\_cooling\_tools} by \cite{2008MNRAS.385.1443S,2017MNRAS.466.2217S} to produce new ion tables to replace "hm2012\_hr.h5" in the Trident configuration file. The steps followed to produce the UV background are:

\begin{itemize}
    \item Choosing a Spectral Energy Distribution (SED) from those produced by STARBURST99 that best represents the stellar population near the simulated atomic cloud. We chose a SED assuming solar metallicity and 3-Myr-old starburst, in agreement with \cite{2013ApJ...770L...4M} concerning the observation of atomic hydrogen in the Galactic centre of our Galaxy. SEDs have wavelengths in $\AA$ on the X-axis and luminosity in $\rm erg \; s^{-1}$ \AA$^{-1}$ on the Y-axis.
    
    \item We converted the brightnesses to specific intensities and produced an \texttt{.out} file with a customised format, so that it can be read by CLOUDY. The Python script that handles this procedure is \texttt{sb99\_cloudy\_interface.py}, and the wavelengths with relative luminosities given as input have been converted to $E$ [Ryd] and $log(J_{\mu})$. We also converted the SEDs representing the \cite{2012ApJ...746..125H} profile provided by CLOUDY and TRIDENT into the same units to evaluate more clearly how important it is to consider a UV background which takes into account star formation. The SED produced with STARBURST99 has a flux at least one order of magnitude higher than the others, in the wavelength range that we are interested in. Our results also appear to be in agreement with \cite{2014ApJ...792....8W}. 
    
    \item Then, we ran a CLOUDY simulation in parallel using 16 cores with the \texttt{CIAOLoop} code, located inside \texttt{Cloudy\_cooling\_tools}, on SUPERMUC-NG in order to obtain ion tables that take into account radiation from stellar feedback. The simulation receives as input the \texttt{.out} file produced in the previous step and was run with a temperature range from $1$ to $10^9$ $\rm K$, hydrogen number density of $10^{-9}$ cm$^{-3}-10^4$ cm$^{-3}$ with step size of 0.0125 dex and at redshift = 0 (as advised in \citealt{2017ApJ...847...59H} and \citealt{2018ApJ...864...96C}).
    
    \item Finally, by exploiting the \texttt{convert\_ion\_balance\_tables} function imported from \texttt{cloudy\_grids} inside \texttt{Cloudy\_cooling\_tools} we generated the ion field output file in \texttt{.h5} format, which we then placed inside the \texttt{.trident} folder as the new default background.
\end{itemize}

\section{Preliminary Results and Conclusions}

\begin{figure}[h!]
   \centering
   \includegraphics[width=1.0\textwidth, angle = 0]{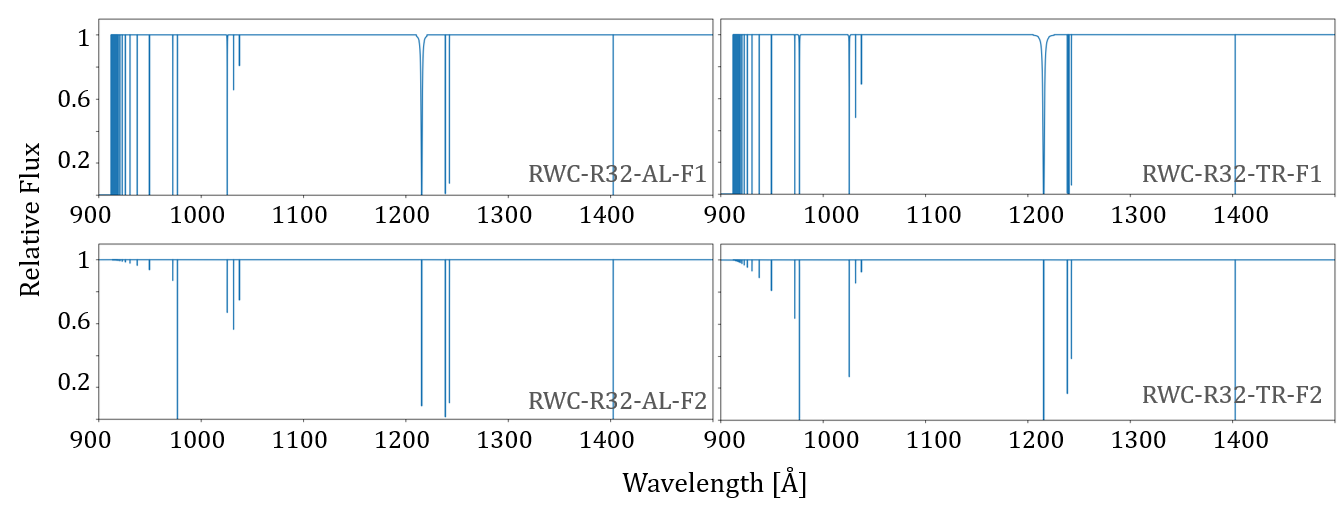}
   \caption{Synthetic spectra of the single cloud models shown in Figure 1. The $4$ spectra were obtained at the same time as the images in Figure 1, where the top two panels correspond to models with cooling floor at $50$ K, while the bottom two panels correspond to models with $10^4$ K as cooling floor, and are all derived imposing zero noise. The lines in absorption represent the following ions: H\,{\sc i}, O\,{\sc vi}, C\,{\sc iii}, C\,{\sc iv}, Si\,{\sc iv}, N\,{\sc v}, Ne\,{\sc viii}, and Mg\,{\sc ii}.}
   \label{spectra}
\end{figure}

The resulting down-the-barrel spectra obtained by following the framework described above can be viewed in Fig. \ref{spectra}. The top two panels correspond to models with a cooling floor at $50$ K, while the bottom two panels correspond to models with $10^4$ K as the cooling floor. By comparing different models representing the interaction of a single atomic cloud with galactic winds, each of them with a different cooling floor and different orientation of the magnetic field, it is possible to see how both phenomena influence the morphology of the atomic cloud and potentially the spectra. We find that the morphological variations of the clouds, which we mentioned earlier, are also reflected in the resulting spectra.\par

While we are currently working on quantifying how much the cooling floor and magnetic fields affect the column density of certain ions such as O\,{\sc vi}, some qualitative features of our spectra already draw our attention. For instance, Figure \ref{spectra} reveals that changing the orientation of the magnetic field in the top two models does not cause large variations in the spectra; while in the bottom two panels the depth of some lines varies depending on the orientation of the magnetic field. Moreover, quenching radiative cooling at higher temperatures has the main effect of reducing the width of the Ly$\alpha$ absorption line.\par

As briefly presented here, the overall aim of our project is to create a framework able to produce synthetic spectra from galactic wind simulations. For the code development part, the PLUTO-TRIDENT interface is now ready to use. The advantage of this tool is that it is quite flexible and can be used to study the physics and chemistry of different astrophysical systems on different scales. In addition, it makes it possible to produce column density maps for individual ions and spectra with customised UV backgrounds that take into account the high star formation that occurs in a starburst galaxy. A further future goal is to produce synthetic spectra simulating observations with different instruments.\par

The authors gratefully acknowledge the Gauss Centre for Supercomputing e.V. (\url{www.gauss-centre.eu}) for funding this project by providing computing time (via grant pn34qu) on the GCS Supercomputer SuperMUC-NG at the Leibniz Supercomputing Centre (\url{www.lrz.de}). WEBB is supported by the Deutsche Forschungsgemeinschaft (DFG) via grant BR2026/25, and by the National Secretariat of Higher Education, Science, Technology, and Innovation of Ecuador, SENESCYT. We also thank the referee for their helpful feedback on our paper.

\bibliographystyle{iaulike}
\bibliography{bibliography}

\section*{Discussion}

\textbf{Q: In the density-temperature plot, the majority of the gas is placed in a line which reflects the adiabatic expansion of the cloud. However, there is some hot and low-density gas that does not follow the adiabatic relation; why?}\\

A:  Shock heating causes the adiabatic expansion of the cloud and produces the hot gas we see on the $n_H$ vs. $T$ maps. There is also radiative heating in these models, but this plays a smaller role in heating up hot gas.\\

\textbf{Q: Do you think relative motions can affect the evolution of the clouds?}\\

A:  Adding relative motion to these models would not be needed because we can always change the reference frame of the simulation (via a Galilean transformation) into one where the cloud is not moving at t=0. To study the interaction of moving/evolving clouds with winds, wind-filament and shock-multicloud models are more adequate than single-cloud models.\\

\textbf{Q: Have you already compared your theoretical spectra with some observed within our Galaxy, in particular the absorption lines of some ions such as O\,{\sc vi}?}\\

A: Studying O\,{\sc vi} in our Galaxy is definitely something we can do in future work, there are also other ions in the Milky Way (e.g. \citealt{2017A&A...607A..48R}).

\end{document}